# A New Approach to Adaptive Signal Processing

**Muhammad Ali Raza Anjum**
Department of Electrical Engineering,
Army Public College of Management and Sciences, Rawalpindi, PAKISTAN
ali.raza.anjum@apcoms.edu.pk

**ABSTRACT**
A unified linear algebraic approach to adaptive signal processing (ASP) is presented. Starting from just Ax=b, key ASP algorithms are derived in a simple, systematic, and integrated manner without requiring any background knowledge to the field. Algorithms covered are Steepest Descent, LMS, Normalized LMS, Kaczmarz, Affine Projection, RLS, Kalman filter, and MMSE/Least Square Wiener filters. By following this approach, readers will discover a synthesis; they will learn that one and only one equation is involved in all these algorithms. They will also learn that this one equation forms the basis of more advanced algorithms like reduced rank adaptive filters, extended Kalman filter, particle filters, multigrid methods, preconditioning methods, Krylov subspace methods and conjugate gradients. This will enable them to enter many sophisticated realms of modern research and development. Eventually, this one equation will not only become their passport to ASP but also to many highly specialized areas of computational science and engineering.

## 1. INTRODUCTION

In last few decades, Adaptive Signal Processing (ASP) has emerged into one of the most prolific areas in Electrical and Computer Engineering (ECE). It is based on the concept of intelligent systems that can automatically adapt to changes in their environment without the need for manual intervention [1]. ASP is gaining popularity with each passing day and is forming the basis of many key future technologies: including robots, gyroscopes, power systems, e-health systems, communication networks, audio and video technologies, etc [2-14]. As a result, ASP has become very important from both practical and pedagogical viewpoints. Therefore, its importance can hardly be over-emphasized.

But before one can enter into the realm of ASP, there are a lot of problems to face. To begin with, ASP lacks a unified framework. Numerous approaches to ASP can be found in literature and each of them tackles ASP in its own particular way. Secondly, the notation is mostly author specific. This makes it very difficult for the reader to understand even a single concept from two different manuscripts. Thirdly, almost every author leaves something out. For example, most authors drop the Kalman filter which has been recognized as one of most crucial topics in postgraduate research. Fourthly, authors make no attempt at making the concepts portable. Finally, the subject has a heavy probabilistic/statistical outlook which is overemphasized most of the time.

This work will provide a unified linear algebraic approach to ASP. Starting from just one equation in one unknown and one observation, all the key ASP algorithms - from Steepest Descent to Kalman filter - will be derived. During the derivation process, notation of algorithms will be kept uniform to make them consistent. Transitions from one algorithm to other will be kept systematic to make them portable. Probability and statistics shall not be invoked to make the algorithms accessible as well. The treatment of ASP will be entirely linear algebraic. Moreover, connection of ASP to other highly specialized domains like extended Kalman filter, reduced-rank adaptive filters, particle filters, multigrid methods, pre-conditioning methods, and Krylov subspace methods will also be established.

## 2. SYSTEM MODEL

Let a system of linear equations be described as,

$$A_m x = b_m \tag{1}$$

With,

$$A_m = \begin{bmatrix} a^T{}_1 \\ a^T{}_2 \\ \vdots \\ a^T{}_m \end{bmatrix} \qquad x = \begin{bmatrix} x_1 \\ x_2 \\ \vdots \\ x_n \end{bmatrix} \qquad b_m = \begin{bmatrix} b_1 \\ b_2 \\ \vdots \\ b_m \end{bmatrix}$$

$x$ represents the unknown system parameters. $A_m$ contains the rows of the input data $a_i$'s that are applied to this system. $b_m$ comprises of the observations $b_i$'s at the output of the system. $x$ is an ($n \times 1$) vector. $A_m$ is an ($m \times n$) matrix. $b_m$ is an ($m \times 1$) vector. According to this nomenclature, $m$ inputs have been applied to the system with $n$ unknown parameters and so far $m$ outputs have been observed.

### 3. LEAST MEAN SQUARES (LMS) ALGORITHM

Beginning with $m = 1$, there is one equation and one observation $b_1$.

$$a_1{}^T x = b_1 \tag{2}$$

Eq. (2) seeks a vector $x$ that has a projection of magnitude $b_1$ over the vector $a_1$. We begin with an arbitrary guess, say $x[k]$ at time-step $k$. Since our guess is purely arbitrary, the projection of $x[k]$ over $a_1$ may not be equal to $b_1$, i.e., $a_1{}^T x[k] \neq b_1$. There will be a difference. This difference can be denoted by an error $e[k]$.

$$e[k] = b_1 - a_1{}^T x[k] \tag{3}$$

$e[k]$ is the error in projection of $x[k]$ over $a_1$. It lies in the direction of $a_1$. The complete error vector $e[k]$ will be $e[k]a_1$. This error can be added to the initial guess $x[k]$ to rectify it. But the question is how much of it to add? Let there be a scalar $\mu$ in the range of $[0,1]$ such that the error vector is multiplied with $\mu$ before adding it to $x[k]$. A value of $\mu = 0$ indicates that no correction is required and the initial guess is accurate whereas a value of $\mu = 1$ indicates that a full correction is required.

$$x[k+1] = x[k] + \mu e[k]a_1 \tag{4}$$

$x[k+1]$ is the updated vector. Substituting Eq. (3) in Eq. (4),

$$x[k+1] = x[k] + \mu(b_1 - a_1{}^T x[k])a_1 \tag{5}$$

If $k$ observations have been received up till the time-step,

$$x[k+1] = x[k] + \mu(b_k - a_k{}^T x[k])a_k \tag{6}$$

Eq. (6) represents the famous LMS algorithm. The parameter $\mu$ in Eq. (6) is known as the step-size. It plays a crucial role in the convergence of the LMS algorithm. A very small step-size makes the algorithm crawl towards the true solution and hence terribly slows it down. This phenomenon is known as *lagging*. A large step-size makes the algorithm leap towards the true solution. Though it will make the algorithm much faster, movement in large steps never allows the algorithm to approach the true solution within a close margin. The algorithm keeps jumping around the true value but never converges to it and a further decrease in error becomes impossible. This phenomenon is known as *misadjustment* [1]. Effect of these phenomenons on the convergence properties of LMS algorithm are illustrated in Figure 1. These effects arise due to the fact that the step-size of LMS algorithm has to be adjusted manually. This is its major drawback. But inherent simplicity of LMS algorithm still keeps it much popular.

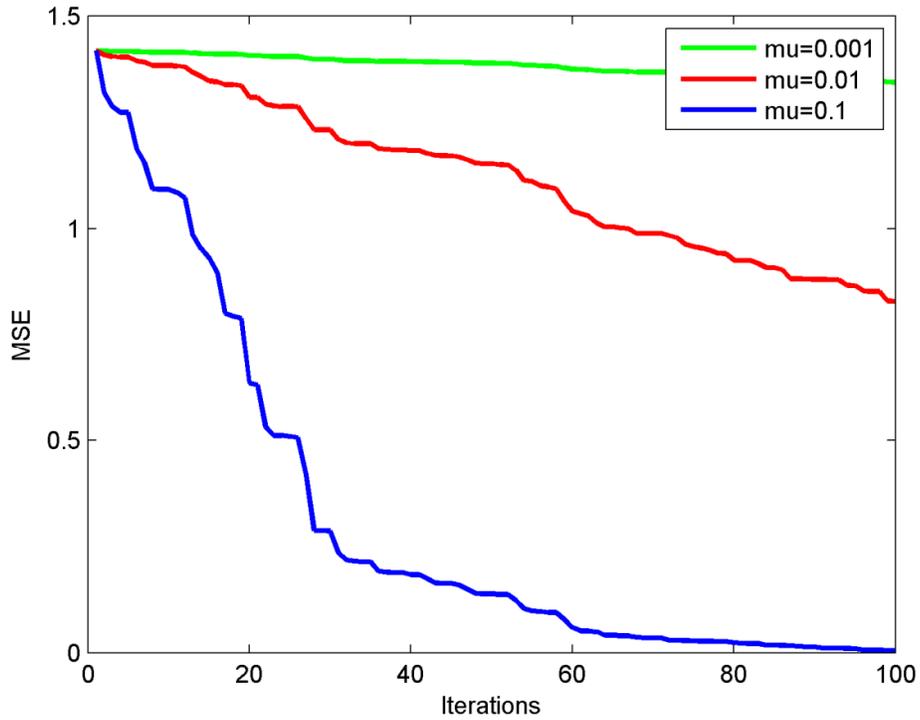

Figure 1. Comparison of the convergence properties of LMS algorithm with different step-sizes. LMS was employed to identify a system with an impulse response of length $n = 5$.

## 4. NORMALIZED LEAST MEAN SQUARES (NLMS) ALGORITHM

NLMS provides an automatic adjustment in step-size. It is based on the criteria of selecting the best step-size for a given iteration. The term best is explained as follows. If error during the iteration is large, step-size is kept large so that the algorithm can quickly catch up with true solution. If the error decreases, step-size is lowered to allow the algorithm to zoom into the true solution. Hence, NLMS tries to select a step-size that minimizes the error in each iteration. In order to show how NLMS achieves it, we have to re-consider Eq. (2). It represents the projection of vector $\boldsymbol{x}$ over the vector $\boldsymbol{a_1}$ such that this projection has magnitude equal to $b_1$. But this is not the usual definition of dot product that represents orthogonal projections [15]. Actual definition requires the normalization of $\boldsymbol{a_1}$ in Eq. (2).

$$\frac{\boldsymbol{a_1}^T \boldsymbol{x}}{\boldsymbol{a_1}^T \boldsymbol{a_1}} = \frac{b_1}{\boldsymbol{a_1}^T \boldsymbol{a_1}} \qquad (7)$$

Orthogonal projections make sure that error is orthogonal and hence, minimum. This in turn appears as a constraint on the step-size as we show now. Continuing in the similar fashion by choosing an arbitrary vector $\boldsymbol{x[k]}$, the projection may not necessarily equal the right hand side and there will again be a difference.

$$\frac{b_1}{\boldsymbol{a_1}^T \boldsymbol{a_1}} - \frac{\boldsymbol{a_1}^T \boldsymbol{x[k]}}{\boldsymbol{a_1}^T \boldsymbol{a_1}} = \frac{1}{\boldsymbol{a_1}^T \boldsymbol{a_1}} (b_1 - \boldsymbol{a_1}^T \boldsymbol{x[k]}) = \frac{1}{\boldsymbol{a_1}^T \boldsymbol{a_1}} e[k]$$

Where,

$$e[k] = (b_1 - \boldsymbol{a_1}^T \boldsymbol{x[k]}) \qquad (8)$$

The error vector will be,

$$\boldsymbol{e[k]} = \frac{1}{\boldsymbol{a_1}^T \boldsymbol{a_1}} e[k] \boldsymbol{a_1} \qquad (9)$$

Adding this correction to the original guess in order to improve it for the next iteration $x[k+1]$,

$$x[k+1] = x[k] + \frac{1}{a_1{}^T a_1} e[k] a_1 \tag{10}$$

Substituting Eq. (8) in Eq. (10),

$$x[k+1] = x[k] + \frac{1}{a_1{}^T a_1}(b_1 - a_1{}^T x[k]) a_1 \tag{11}$$

Or in general if $k$ observations have been received till the time-step,

$$x[k+1] = x[k] + \frac{1}{a_k{}^T a_k}(b_k - a_k{}^T x[k]) a_k \tag{12}$$

Eq. (12) represents the NLMS algorithm. Comparing Eq. (12) with Eq. (6), it can be observed that,

$$\mu = \frac{1}{a_k{}^T a_k} \tag{13}$$

In contrast to the LMS algorithm, step size of the NLMS algorithm is a variable which is automatically adjusted for each input row of data $a_k$ according to its norm $a_k{}^T a_k$. By this automatic adjustment of step size, NLMS is able to avoid the problems of lagging and misadjustment that plague the LMS algorithm. Therefore, it has better convergence properties than LMS algorithm. Since NLMS achieves this advantage by the normalization step performed in Eq. (7), hence follows the name normalized LMS [1].

## 5. KACZMARZ ALGORITHM

Eq. (11) for NLMS algorithm can be re-written as,

$$x[k+1] = x[k] + \frac{1}{a_k{}^T a_k}(b_k - a_k{}^T x[k]) a_k \tag{Repeat}$$

This is the Kaczmarz equation. Kaczmarz algorithm was originally proposed for solving the under-determined systems ($m < n$) [16]. Due to a limited number of rows, the Kaczmarz equation keeps jumping back to the first row after it has reached the last row until the solution converges. Hence, Kaczmarz algorithm is recurrent in terms of rows. Whereas in NLMS, new rows are added continuously due to the arrival of new data and the system, therefore, becomes over-determined, i.e., more rows than the columns. In this case, previous rows are never used. Therefore, NLMS algorithm is not recurrent. Otherwise both algorithms are identical.

## 6. AFFINE PROJECTION (AP) ALGORITHM

Re-writing Eq. (11) for NLMS algorithm,

$$x[k+1] = x[k] + a_1(a_1{}^T a_1)^{-1}(b_1 - a_1{}^T x[k]) \tag{14}$$

NLMS algorithm tries to reduce the error in Eq. (3) with respect to a single row of data $a_1$. The idea behind the AP algorithm is to choose a step-size that minimizes the error with respect to all the $k$ rows of data that have been received up to time-step $k$ to improve its convergence properties [17]. Hence, the vector $a_1$ in Eq. (14) is replaced by a matrix $A_k$ which contains all the $k$ data rows.

$$x[k+1] = x[k] + A_k{}^T (A_k A_k{}^T)^{-1}(b_k - A_k x[k]) \tag{15}$$

As long as the number of these data rows remains less than dimension of the system ($k < n$), the system can be solved by Eq. (15). Such a system is called an under-determined system [18]. It means that there are more unknowns than the number of equations which in turn implies that much less data is available. Eq. (15) is known as AP algorithm and $A_k{}^T (A_k A_k{}^T)^{-1}$ is defined as the pseudoinverse for an under-determined system [18]. It is important to note that the pseudoinverse $A_k{}^T(A_kA_k{}^T)^{-1}$ is dependent on time-step $k$. Thus, the solution to Eq. (15) is obtained by computing the term $A_k{}^T(A_kA_k{}^T)^{-1}$ at each time-step. In this way, AP algorithm may appear much more complex than NLMS and Kaczmarz algorithms. Also there is no indirect way of computing the pseudoinverse. However, these drawbacks are offset by its much faster convergence as compared to NLMS and Kaczmarz algorithms.

## 7. RECURSIVE LEAST SQUARES (RLS) ALGORITHM

AP algorithm implies that the amount of data available is much less for the system to be full-determined ($k < n$). However, this case seldom occurs in practice. On the contrary, more data keeps arriving with each time-step and the number of data rows exceeds the dimensions of the system ($k > n$). As a result, the system becomes over-determined. There are more equations than the unknowns. In this case, the pseudoinverse for an under-determined systems $A_k{}^T(A_kA_k{}^T)^{-1}$ in Eq. (15) is be replaced by the Least Squares pseudoinverse defined for an over-determined case $(A_k{}^TA_k)^{-1}A_k{}^T$ [18].

$$x[k+1] = x[k] + (A_k{}^TA_k)^{-1}A_k{}^T(b_k - A_kx[k])$$ (16)

But before we can proceed with the iterative solution, we must consider $(A_k{}^TA_k)^{-1}$ term in Eq. (16). This term will make Eq. (16) converge in one iteration.

$$x[k+1] = x[k] + (A_k{}^TA_k)^{-1}A_k{}^Tb_k - (A_k{}^TA_k)^{-1}A_k{}^TA_kx[k] = x[k] + x - x[k]$$ (17)

Hence,

$$x[k+1] = x$$ (18)

But if we can compute this term beforehand, then the whole point of iterative solution becomes useless and the system can be directly solved in one step using Least Squares. Luckily, a lemma is available that can avoid the direct computation of $(A_k{}^TA_k)^{-1}$. This lemma is known as the *matrix inversion lemma* [18]. Here, we explain the great advantage achieved by this lemma over the AP algorithm where no such flexibility is available. We begin by examining the $A_k{}^TA_k$ term. This term can be decomposed as a sum of $k$ rank-1 matrices.

$$A_k{}^TA_k = [a_1 \quad a_2 \quad \dots \quad a_k]\begin{bmatrix} a^T{}_1 \\ a^T{}_2 \\ \vdots \\ a^T{}_k \end{bmatrix} = a_1a^T{}_1 + a_2a^T{}_2 + \dots + a_{k-1}a^T{}_{k-1} + a_ka^T{}_k$$

$$A_k{}^TA_k = A_{k-1}{}^TA_{k-1} + a_ka^T{}_k$$ (19)

The term $a_ka^T{}_k$ is known as the rank-1 *update-term* because it updates the $A_{k-1}{}^TA_{k-1}$ when a single new row of data $a_k$ is added to the system at time-step $k$. The matrix inversion lemma computes the inverse $(A_k{}^TA_k)^{-1}$ by incorporating the rank-1 update $a_ka^T{}_k$ into the inverse $(A_{k-1}{}^TA_{k-1})^{-1}$. An identity matrix can be chosen as a starting candidate for $(A_{k-1}{}^TA_{k-1})^{-1}$ term.

$$(A_{k-1}{}^TA_{k-1} + a_ka_k{}^T)^{-1} = (A_{k-1}{}^TA_{k-1})^{-1} - \frac{(A_{k-1}{}^TA_{k-1})^{-1}a_ka_k{}^T(A_{k-1}{}^TA_{k-1})^{-1}}{(1 + a_k{}^T(A_{k-1}{}^TA_{k-1})^{-1}a_k)}$$ (20)

In this way, the system does not have to wait for all the rows of data to form its estimate $x$. Instead, the estimate is continually updated with the arrival of every new row of data. Also, the lemma seamlessly updates the estimate with every new observation and saves the toil of solving the system all over when new data arrives. Substituting Eq. (19) in Eq. (16) and modifying Eq. (16) for one row of data at a time,

$$x[k+1] = x[k] + \left(A_{k-1}{}^T A_{k-1} + a_k a^T{}_k\right)^{-1} a_k(b_k - a^T{}_k x[k]) \tag{21}$$

Eq. (21) is known as the RLS algorithm. Due to its dependency on matrix inversion lemma, RLS is much more complex than the rest of the algorithms discussed so far. But despite that, it has been the most popular algorithm to date. Reasons for this are its ease of implementation, its excellent convergence properties, and its ability to update the estimate after every new observation without the need of solving the entire system of linear equations all over again. Figure 2 depicts the superior convergence properties of RLS algorithm as compared to the NLMS and LMS algorithms. Table 1 compares the computational complexity of the algorithms discussed so far [17].

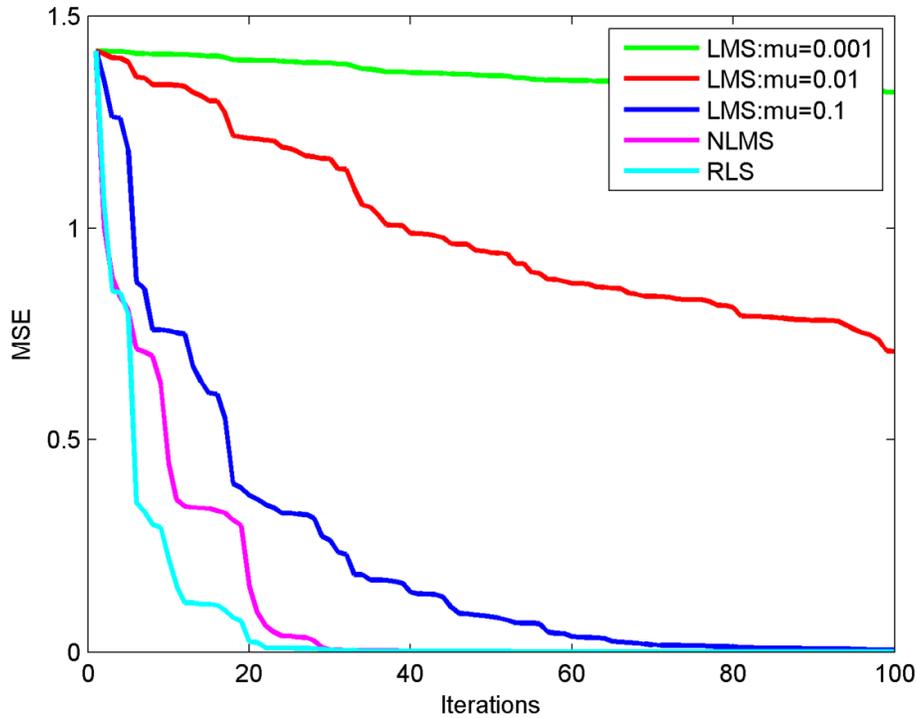

Figure 2. Comparison of the convergence properties of RLS, NLMS, and LMS algorithms. These algorithms were employed to identify a time-invariant system with an impulse response of length $n = 5$. System was over-determined.

Table 1. Comparison of the computational complexity of various ASP algorithms

| Algorithm | Complexity |
|-----------|------------|
| LMS | $2N + 1$ |
| NLMS | $3M + 2$ |
| Kaczmarz | $3M + 2$ |
| AF | $N^2$ |
| RLS | $N^2$ |

## 8. KALMAN FILTER

All the algorithms that have been discussed up till now have a major underlying assumption. They provide an iterative solution to a time-invariant system. By time-invariant we mean that the vector $x$ depicting the system parameters in Eq. (1) remains unchanged during the iterative process. This assumption is relatively weak as it often happens in physical situations that the parameters of a system change during the convergence process, say for example in wireless communications [19]. Therefore, an algorithm must track the system during the transition process in addition to assuring the convergence within the intervals between the transitions. These are precisely the objectives of Kalman filter [20]. In order to show that how Kalman filter accomplishes it, we modify the system model in Eq. (1) to incorporate the changes in nomenclature that arise due to the time varying nature of $\boldsymbol{x}$.

$$\boldsymbol{A_{mn}x_n} = \boldsymbol{b_m} \tag{22}$$

$\boldsymbol{A_{mn}}$ is an $(m \times n)$ matrix. $\boldsymbol{x}$ is an $(n \times 1)$ vector. $\boldsymbol{b_m}$ is an $(m \times 1)$ vector. The system is over-determined $(m > n)$. $m$ data rows and $m$ observations have been received and the solution has been obtained using RLS algorithm in Eq. (21). Let us assume that the system in Eq. (22) changes its state from $\boldsymbol{x_n}$ to $\boldsymbol{\hat{x}_n}$. This change of state will create a new set of unknowns $\boldsymbol{\hat{x}_n}$ with the following linear relationship to previous unknowns $\boldsymbol{x_n}$.

$$\boldsymbol{\hat{x}_n} = \boldsymbol{F_{nn}x_n} + \boldsymbol{c_n} \tag{23}$$

$\boldsymbol{\hat{x}_n}$ is an $(n \times 1)$ vector with $n$ new system unknowns. $\boldsymbol{F}$ is an $(n \times n)$ matrix responsible for the change of state. It is also known as the *state-transition matrix*. $\boldsymbol{c_n}$ is an $(n \times 1)$ vector of constants. Re-writing Eq. (23),

$$-\boldsymbol{F_{nn}x_n} + \boldsymbol{\hat{x}_n} = \boldsymbol{c_n} \tag{24}$$

Combining Eq. (22) and (24) yields,

$$\begin{bmatrix} \boldsymbol{A_{mn}} & \boldsymbol{0_{nn}} \\ -\boldsymbol{F_{nn}} & \boldsymbol{I_{nn}} \end{bmatrix} \begin{bmatrix} \boldsymbol{x_n} \\ \boldsymbol{\hat{x}_n} \end{bmatrix} = \begin{bmatrix} \boldsymbol{b_m} \\ \boldsymbol{c_n} \end{bmatrix} \tag{25}$$

$\boldsymbol{0_{nn}}$ is an $(n \times n)$ matrix with all zero entries. $\boldsymbol{I_{nn}}$ is $(n \times n)$ identity matrix. Least Squares solution to Eq. (25) is,

$$\begin{bmatrix} \boldsymbol{A^T_{mn}} & -\boldsymbol{F^T_{nn}} \\ \boldsymbol{0_{nn}} & \boldsymbol{I_{nn}} \end{bmatrix} \begin{bmatrix} \boldsymbol{A_{mn}} & \boldsymbol{0_{nn}} \\ -\boldsymbol{F_{nn}} & \boldsymbol{I_{nn}} \end{bmatrix} \begin{bmatrix} \boldsymbol{x_n} \\ \boldsymbol{\hat{x}_n} \end{bmatrix} = \begin{bmatrix} \boldsymbol{A^T_{mn}} & -\boldsymbol{F^T_{nn}} \\ \boldsymbol{0_{nn}} & \boldsymbol{I_{nn}} \end{bmatrix} \begin{bmatrix} \boldsymbol{b_m} \\ \boldsymbol{c_n} \end{bmatrix} \tag{26}$$

Adopting column-wise multiplication in Eq. (26),

$$\left[ \begin{bmatrix} \boldsymbol{A^T_{mn}} \\ \boldsymbol{0_{nn}} \end{bmatrix} \begin{bmatrix} \boldsymbol{A_{mn}} & \boldsymbol{0_{nn}} \end{bmatrix} + \begin{bmatrix} -\boldsymbol{F^T_{nn}} \\ \boldsymbol{I_{nn}} \end{bmatrix} \begin{bmatrix} -\boldsymbol{F_{nn}} & \boldsymbol{I_{nn}} \end{bmatrix} \right] \begin{bmatrix} \boldsymbol{x_n} \\ \boldsymbol{\hat{x}_n} \end{bmatrix} = \begin{bmatrix} \boldsymbol{A^T_{mn}} \\ \boldsymbol{0_{nn}} \end{bmatrix} \boldsymbol{b_m} + \begin{bmatrix} -\boldsymbol{F^T_{nn}} \\ \boldsymbol{I_{nn}} \end{bmatrix} \boldsymbol{c_n} \tag{27}$$

In order to find the value of new unknowns, we only need the inverse of the first term in the square brackets on the left hand side of Eq. (27).

$$\begin{bmatrix} \boldsymbol{A^T_{mn}} \\ \boldsymbol{0_{nn}} \end{bmatrix} \begin{bmatrix} \boldsymbol{A_{mn}} & \boldsymbol{0_{nn}} \end{bmatrix} + \begin{bmatrix} -\boldsymbol{F^T_{nn}} \\ \boldsymbol{I_{nn}} \end{bmatrix} \begin{bmatrix} -\boldsymbol{F_{nn}} & \boldsymbol{I_{nn}} \end{bmatrix} \tag{28}$$

But we do not have to compute this inverse from right from start. From Eq. (25), we observe that $n$ new rows and $n$ new columns have been added to the system in Eq. (22) by $\boldsymbol{F_{nn}}$ and $\boldsymbol{I_{nn}}$ matrices. These extra rows and columns can be incorporated as rank-1 updates in the original matrix $\boldsymbol{A_{mn}}$. This is because the zeros matrix $\boldsymbol{0_{nn}}$ to the right of $\boldsymbol{A_{mn}}$ in Eq. (25) has no impact on the rows of $\boldsymbol{A_{mn}}$ other than increasing the length of individual rows from $n$ to $2n$ by appendage $n$ zeros at the end. This also leaves the product $\boldsymbol{A^T_{mn}A_{mn}}$ unchanged. Only its dimensions have increased from $n \times n$ to $2n \times 2n$ with all the entries zeros expect the first $n \times n$ ones which are equal to $\boldsymbol{A^T_{mn}A_{mn}}$.

$$\begin{bmatrix} \boldsymbol{A^T_{mn}} \\ \boldsymbol{0_{nn}} \end{bmatrix} \begin{bmatrix} \boldsymbol{A_{mn}} & \boldsymbol{0_{nn}} \end{bmatrix} = \begin{bmatrix} \boldsymbol{A^T_{mn}A_{mn}} & \boldsymbol{0_{nn}} \\ \boldsymbol{0_{nn}} & \boldsymbol{0_{nn}} \end{bmatrix} \tag{29}$$

Therefore, only the change in nomenclature is required to be precise. Otherwise the rest stays the same with this term. Therefore, we term the vector $\begin{bmatrix} \boldsymbol{A_{mn}} & \boldsymbol{0} \end{bmatrix}$ to $\boldsymbol{A_{m(2n)}}$, meaning that $\boldsymbol{A_{m(2n)}}$ has $m$ rows each of length $2n$ with last $n$ entries all zero. Similarly, we term $\begin{bmatrix} -\boldsymbol{F_{nn}} & \boldsymbol{I_{nn}} \end{bmatrix}$ as $\boldsymbol{\hat{F}_{n(2n)}}$ because each of its rows is subjoined by the corresponding row of the identity matrix $\boldsymbol{I_{nn}}$. Both the new and old unknowns in Eq. (25) are jointly represented by $\boldsymbol{x_{2n}}$. Eq. (25) becomes,

$$\begin{bmatrix} \boldsymbol{A_{m(2n)}} \\ \boldsymbol{\hat{F}_{n(2n)}} \end{bmatrix} \boldsymbol{x_{2n}} = \begin{bmatrix} \boldsymbol{b_m} \\ \boldsymbol{c_n} \end{bmatrix} \tag{30}$$

Subsequently, Eq. (28) becomes,

$$\begin{bmatrix} A^T{}_{mn} \\ 0_{nn} \end{bmatrix} [A_{mn} \quad 0_{nn}] + \begin{bmatrix} -F^T{}_{nn} \\ I_{nn} \end{bmatrix} [-F_{nn} \quad I_{nn}] = A^T{}_{m(2n)} A_{m(2n)} + \widehat{F}^T{}_{n(2n)} \widehat{F}_{n(2n)} \tag{31}$$

Matrix $\widehat{F}^T{}_{n(2n)} \widehat{F}_{n(2n)}$ in Eq. (31) can be decomposed into a series of rank-one updates,

$$A^T{}_{m(2n)} A_{m(2n)} + \widehat{F}^T{}_{n(2n)} \widehat{F}_{n(2n)} = A^T{}_{m(2n)} A_{m(2n)} + \sum_{i=0}^{n} \widehat{f}_{i(2n)} \widehat{f}_{i(2n)}{}^T \tag{32}$$

So the inverse required in Eq. (26) can be written as,

$$\left[ \begin{bmatrix} A^T{}_{mn} \\ 0_{nn} \end{bmatrix} [A_{mn} \quad 0_{nn}] + \begin{bmatrix} -F^T{}_{nn} \\ I_{nn} \end{bmatrix} [-F_{nn} \quad I_{nn}] \right]^{-1} = \left( A^T{}_{m(2n)} A_{m(2n)} + \sum_{i=0}^{n} \widehat{f}_{i(2n)} \widehat{f}_{i(2n)}{}^T \right)^{-1} \tag{33}$$

The rank-1 updates of $\widehat{F}_{n(2n)}$ matrix can be used to form the inverse in Eq. (33) using matrix inversion lemma. This inverse can then be used to from the pseudoinverse to compute the value of $x_{2n}$.

$$x_{2n} = \left( A^T{}_{m(2n)} A_{m(2n)} + \widehat{F}^T{}_{n(2n)} \widehat{F}_{n(2n)} \right)^{-1} \left( A^T{}_{m(2n)} b_m + \widehat{F}^T{}_{n(2n)} c_n \right) \tag{34}$$

In this way, the values of the new unknowns can be determined without the need of explicitly computing the pseudoinverse. It is important to note that despite the system has changed state, new data measurements have not yet arrived. By solving for $x_{2n}$ in advance, values of new unknowns can be *predicted* and the values of previous unknowns can be *smoothed*. Hence, Eq. (34) is known as *prediction equation*. When the new measurement arrives, it is added as a new row of data to Eq. (30).

$$\begin{bmatrix} A_{m(2n)} \\ \widehat{F}_{n(2n)} \\ a^T{}_{k(2n)} \end{bmatrix} x_{2n} = \begin{bmatrix} b_m \\ c_n \\ b_k \end{bmatrix} \tag{35}$$

RLS can now be invoked to update the predicted vector $x_{2n}[k]$ in the light of new measurements.

$$x_{2n}[k+1] = x_{2n}[k] + \left( A^T{}_{m(2n)} A_{m(2n)} + \widehat{F}^T{}_{n(2n)} \widehat{F}_{n(2n)} + a_{k(2n)} a^T{}_{k(2n)} \right)^{-1} a_{k(2n)} \left( b_k - a^T{}_{k(2n)} x_{2n}[k] \right) \tag{36}$$

Again the matrix inversion lemma can be used to compute the term $\left( A^T{}_{m(2n)} A_{m(2n)} + \widehat{F}^T{}_{n(2n)} \widehat{F}_{n(2n)} + a_{k(2n)} a^T{}_{k(2n)} \right)^{-1}$ by employing the rank-1 update $a_{k(2n)} a^T{}_{k(2n)}$ in the term $A^T{}_{m(2n)} A_{m(2n)} + \widehat{F}^T{}_{n(2n)} \widehat{F}_{n(2n)}$ calculated in Eq. (32). Eq. (36) is known as *update equation*. Following term in update equation,

$$\left( A^T{}_{m(2n)} A_{m(2n)} + \widehat{F}^T{}_{n(2n)} \widehat{F}_{n(2n)} + a_{k(2n)} a^T{}_{k(2n)} \right)^{-1} a_{k(2n)} \tag{37}$$

is known as *Kalman gain*. As more and more measurements arrive, Eq. (36) is re-run to update the predicated estimate. In case the state of the system changes again, new unknowns can be added to the system in the same manner as described before in Eq. (25) and the same procedure can be repeated. Though we have kept the information of old states together with the new states by the process of *smoothing*, it is not always necessary to do so. As soon as the measurements arrive and the estimated is update, old estimates can be thrown away altogether to keep the information about the most recent state only. This change of states which arises due to the time-varying nature of the system is the only feature that distinguishes Kalman filter from RLS. Otherwise, Kalman filter is similar to RLS because they perform identically for a time-invariant system. For this reason, Kalman filter is often called the time varying RLS [21].

## 9. STEEPEST DESCENT (SD) ALGORITHM

Eq. (5) for LMS algorithm can be expanded as,

$$x[k+1] = x[k] + \mu \left( a_k b_k - a_k a_k{}^T x[k] \right) \tag{38}$$

Instead of taking a row at a time in Eq. (38), SD is interested in the average of all the $k$ rows that have been received so far.

$$\frac{1}{k}(\boldsymbol{a_1}\boldsymbol{a^T}_1 + \boldsymbol{a_2}\boldsymbol{a^T}_2 + \cdots + \boldsymbol{a_k}\boldsymbol{a^T}_k) = \frac{1}{k}\boldsymbol{a_1}\boldsymbol{a^T}_1 + \frac{1}{k}\boldsymbol{a_2}\boldsymbol{a^T}_2 + \cdots + \frac{1}{k}\boldsymbol{a_k}\boldsymbol{a^T}_k \qquad (39)$$

$\frac{1}{k}$ represents the probability of each row as each row is equally likely.

$$P(\boldsymbol{a_1})\boldsymbol{a_1}\boldsymbol{a^T}_1 + P(\boldsymbol{a_2})\boldsymbol{a_2}\boldsymbol{a^T}_2 + \cdots + P(\boldsymbol{a_k})\boldsymbol{a_k}\boldsymbol{a^T}_k = \sum_{i=1}^{k}P(\boldsymbol{a_i})\boldsymbol{a_i}\boldsymbol{a^T}_i = E\{\boldsymbol{a_i}\boldsymbol{a^T}_i\} \qquad (40)$$

$E$ is an expectation operator and stands for expected or average value. Similarly for $\boldsymbol{a_k}b_k$ term in Eq. (39),

$$\frac{1}{k}\boldsymbol{a_1}b_1 + \frac{1}{k}\boldsymbol{a_2}b_2 + \cdots + \frac{1}{k}\boldsymbol{a_k}b_k = P(\boldsymbol{a_1})\boldsymbol{a_1}b_1 + P(\boldsymbol{a_2})\boldsymbol{a_2}b_2 + \cdots + P(\boldsymbol{a_k})\boldsymbol{a_k}b_k \qquad (41)$$

$$E\{\boldsymbol{a_i}b_i\} = \sum_{i=1}^{k}P(\boldsymbol{a_i})\boldsymbol{a_i}b_i \qquad (42)$$

Substituting Eqs. (42) and (40) in Eq. (38),

$$\boldsymbol{x}[\boldsymbol{k+1}] = \boldsymbol{x}[\boldsymbol{k}] + \mu(E\{\boldsymbol{a_i}b_i\} - E\{\boldsymbol{a_i}\boldsymbol{a^T}_i\}\boldsymbol{x}[\boldsymbol{k}]) \qquad (43)$$

Whereas,

$$\boldsymbol{R} = E\{\boldsymbol{a_i}\boldsymbol{a^T}_i\} \qquad (44)$$
$$\boldsymbol{P} = E\{\boldsymbol{a_i}b_i\} \qquad (45)$$

$\boldsymbol{R}$ is known as the *auto-correlation matrix* and $\boldsymbol{P}$ as the *cross-correlation matrix* [1]. Substituting Eqs. (44) and (45) in Eq. (43),

$$\boldsymbol{x}[\boldsymbol{k+1}] = \boldsymbol{x}[\boldsymbol{k}] + \mu(\boldsymbol{P} - \boldsymbol{Rx}[\boldsymbol{k}]) \qquad (46)$$

Eq. (46) represents the Steepest Descent algorithm. This equation differs from Eq. (6) of LMS algorithm only in terms of $\boldsymbol{R}$ and $\boldsymbol{P}$ matrices. They are formed in SD algorithm by taking the average of all the available data rows as depicted in Eqs. (40) and (42). LMS algorithm, on the other hand, only considers one row of data at a time. It has been demonstrated that if the LMS algorithm is run repeatedly for a given problem, then on the average its performance will be equal to that of LMS algorithm [1]. But $\boldsymbol{R}$ and $\boldsymbol{P}$ matrices are usually not available beforehand and computing them at run-time can be expensive. Therefore, LMS algorithm is much more popular than SD algorithm due to its simplicity.

## 10. MMSE AND LEAST SQUARES WIENER FILTERS
As Eq. (46) represents the recursion equation for SD algorithm, $(\boldsymbol{P} - \boldsymbol{Rx}[\boldsymbol{k}])$ must be the gradient.

$$\nabla = \boldsymbol{P} - \boldsymbol{Rx}[\boldsymbol{k}] \qquad (47)$$

Setting the gradient in Eq. (47) to zero and solving directly for $\boldsymbol{x}$,

$$\boldsymbol{Rx} = \boldsymbol{P} \qquad (48)$$

Eq. (48) is known as Minimum Mean Square Error (MMSE) Wiener-Hop equation. $\boldsymbol{x}$ is now called MMSE Wiener Filter [1]. Expanding Eq. (46),

$$E\{\boldsymbol{a_i}\boldsymbol{a^T}_i\}\boldsymbol{x} = E\{\boldsymbol{a_i}b_i\} \qquad (49)$$

As can be observed from Eqs. (39) and (40),

$$E\{a_i a^T{}_i\} = \sum_{i=1}^{k} P(a_i) a_i a^T{}_i = \frac{1}{k}(a_1 a^T{}_1 + a_2 a^T{}_2 + \cdots + a_k a^T{}_k) = \frac{1}{k} A_k{}^T A_k \tag{50}$$

Similarly from Eqs. (41) and (42),

$$E\{a_i b_i\} = \frac{1}{k}(a_1 b_1 + a_2 b_2 + \cdots + a_k b_k) = \frac{1}{k}[a_1 \quad a_2 \quad \ldots \quad a_k]\begin{bmatrix} b_1 \\ b_2 \\ \vdots \\ b_k \end{bmatrix} = \frac{1}{k} A_k{}^T b_k \tag{51}$$

Substituting Eqs. (50) and (51) in Eq. (48),

$$A_k{}^T A_k x = A_k{}^T b_k \tag{52}$$

Eq. (52) is known as the Least Square Wiener-Hop equation. $x$ is now called Least Squares Wiener Filter [1].

## 11. RELATIONSHIP WITH OTHER AREAS

This section establishes the relationship of the work presented in this article to highly specialized fields that are currently the focus of research and development in ECE and applied mathematics. This will enable the readers to explore further opportunities for research and innovation.

### 11.1. Reduced rank adaptive filters

We begin our discussion of reduced-rank adaptive filters with Eq. (52).

$$x = \left(A_k{}^T A_k\right)^{-1} A_k{}^T b_k \tag{53}$$

Least Squares Wiener filter $x$ in Eq. (52) requires the inversion of $A_k{}^T A_k$ matrix. The matrix $A_k{}^T A_k$ is also known as *covariance matrix*. For it to be invertible, it must have full-rank. But this assumption of full-rank has been strongly questioned by the proponents of reduced rank adaptive filters [22]. They claim that such full-rank is not available often in practice, say for example in sensor array processing. This claim not only puts the invertibility of $A_k{}^T A_k$ matrix in question but also the challenges the entire possibility of finding the LS wiener filter. However, the rank of $A_k{}^T A_k$ can be easily analyzed because it is a symmetric matrix. A symmetric matrix can be decomposed into a set of real eigenvalues and orthogonal eigenvectors [15].

$$A_k{}^T A_k = Q \Lambda Q^T \tag{54}$$

$\Lambda$ is an $(n \times n)$ diagonal matrix of eigenvalues. $Q$ is an $(n \times n)$ matrix of eigenvectors. For a matrix to be of full-rank, none of its eigenvalues should be zero. Even if one of them is zero, the matrix becomes non-invertible. In this case, a Moore and Penrose's pseudoinverse can be formed using Eq. (54) [18]. In order to explain the key idea here, we have to concentrate our attention on $Q$ matrix for a moment. $Q$ is an orthogonal matrix, i.e., $Q^T Q = I$. This implies that inverse of $Q$ is equal to its transpose, i.e., $Q^{-1} = Q^T$. Inverting $A_k{}^T A_k$ would mean,

$$\left(A_k{}^T A_k\right)^{-1} = (Q \Lambda Q^T)^{-1} = (Q^T)^{-1} \Lambda^{-1} Q^{-1} = Q \Lambda^{-1} Q^T \tag{55}$$

Hence, to form the inverse $\left(A_k{}^T A_k\right)^{-1}$ we only need to invert the eigenvalues. But since the matrix $A_k{}^T A_k$ is rank-deficient, at least one of its eigenvalues must be zero. Inverting such an eigenvalue will create problems. Idea behind the Moore and Penrose's pseudoinverse is to leave such an eigenvalue un-inverted while inverting the rest [18].

$$\left(A_k{}^T A_k\right)^+ = Q \Lambda^{-1} Q^T \tag{56}$$

$\left(A_k{}^T A_k\right)^+$ indicates the Moore and Penrose's pseudoinverse. A reduced rank LS Wiener filter can then be obtained by,

$$x = \left(A_k{}^T A_k\right)^+ A_k{}^T b_k \tag{57}$$

Similarly, the argument can be extended to RLS algorithm which also requires the inversion of the covariance matrix.

$$x[k+1] = x[k] + \left(A_{k-1}{}^T A_{k-1} + a_k a^T{}_k\right)^{-1} a_k (b_k - a^T{}_k x[k])$$ (Repeat)

As we know from Eq. (19),

$$A_k{}^T A_k = A_{k-1}{}^T A_{k-1} + a_k a^T{}_k$$ (Repeat)

Therefore,

$$x[k+1] = x[k] + \left(A_k{}^T A_k\right)^{-1} a_k (b_k - a^T{}_k x[k])$$ (58)

Here again, if the rank of covariance matrix is not full, pseudo inverse $\left(A_k{}^T A_k\right)^+$ can be used in place of $\left(A_k{}^T A_k\right)^{-1}$.

$$x[k+1] = x[k] + \left(A_k{}^T A_k\right)^+ a_k (b_k - a^T{}_k x[k])$$ (59)

However, this is not the only case argued by the theory of reduced rank adaptive filters. Now that we can form the pseudoinverse, rank of the covariance matrix can be deliberately reduced if some of its eigenvalues are very small. In this way, larger eigenvalues of the covariance matrix can be given preference which have a larger impact on the system performance. This idea is known by the name of *compressing [22]*. It allows the removal of unnecessary data which has no or very less impact on systems performance. As a result, data processing ability of system is greatly improved.

### 11.2. Extended Kalman filter for non-linear control/signal processing

In non-linear systems, the system model in Eq. (22) and state transition model in Eq. (23) become non-linear. In this case, the system matrix $A_{mn}$ and the state transition matrix $F_{nn}$ for Kalman filter are obtained from a first order Taylor series expansion of these models, a technique called *linearization*. Rest stays the same and Eq. (36) is now called the extended Kalman filter (EKF) [23]. It is the most popular filter for tackling non-linear filtering problems and forms the core of non-linear signal processing, non-linear control theory, and robotics.

### 11.3. Particle filters and Bayesian signal processing

If more terms of the Taylor series are retained during the linearization process, higher order EKF is obtained which performs much better than its predecessor. But these extra terms introduce additional complexity. Hence, there is a classic tradeoff between complexity and accuracy. An alternative approach to this problem is to use Monte Carlo techniques to estimate system parameters for better performance. In this case, system and state transition matrices in Eqs. (22) and (23) are replaced by probability estimates and Eq. (36) is now termed as a particle filter. It is major driving concept behind particle filtering theory and Bayesian signal processing [24].

### 11.4. Multigrid methods for boundary value problems

Repeating Eq. (1) for the system model,

$$A_m x = b_m$$ (Repeat)

If we make an arbitrary guess $x[k]$ for the unknown $x$, then the right hand side may not be equal to the left hand side and there will be a difference. This ensuing difference is known as the residue term $r_m$ in multigrid methods.

$$r_m = b_m - A_m x[k]$$ (60)

Multiplying both sides of Eq. (60) with the psuedoinverse $(A^T{}_m A_m)^{-1} A^T{}_m$,

$$(A^T{}_m A_m)^{-1} A^T{}_m r_m = (A^T{}_m A_m)^{-1} A^T{}_m b_m - (A^T{}_m A_m)^{-1} A^T{}_m A_m x[k]$$ (61)

From Eq. (61), it follows,

$$(A^T{}_m A_m)^{-1} A^T{}_m r_m = x - x[k] = e \tag{62}$$

Or,

$$x = x[k] + e \tag{63}$$

$e$ is the error between the true solution and our guess. It is termed as error in the estimate or simply the error. Re-arranging Eq. (62),

$$A_m e = r_m \tag{64}$$

Eq. (64) represents the relationship between the error and the residue. It is identical in structure to the system model in Eq. (1). Residue $r_m$ takes place of the observations $b_m$ and the error $e$ becomes the unknown. However trivial it may seem, it signifies a very important concept behind multigrid methods. An initial guess $x[k]$ for the estimate $x$ is made. This guess is used to obtain the residue $r_m$ in Eq. (60). The residue can then be used to solve for the error $e$ in Eq. (64). Once Eq. (64) is solved, $e$ can be added back to $x[k]$ to find the true estimate $x$. Now we move on to explain the second driving concept behind the multigrid methods. Eq. (6) for LMS algorithm can be used to solve Eq. (1).

$$x[k+1] = x[k] + \mu(b_k - a_k{}^T x[k]) a_k \tag{Repeat}$$

It can also be used to solve Eq. (62).

$$e[k+1] = e[k] + \mu(r_k - a_k{}^T e[k]) a_k \tag{65}$$

This is because Eq. (64) is similar in structure to Eq. (1). Initially Eq. (6) is run to obtain $x[k]$. Then $x[k]$ is plugged into Eq. (60) to obtain $r_m$. After that Eq. (65) is run to directly improve the error in the estimate $e$. It happens in practice that error constitutes a mixture of high and low frequency oscillations, say for example in the numerical solution of Laplace Equation [18]. High frequency oscillations in the error die out quickly as manifested in fast initial convergence of the algorithm. Low frequency oscillations, on the other hand, tend to linger on. These low frequency errors are responsible for settling the algorithm in the steady state in which even after significant number of iterations, reduction in error is negligible. It is well-known from multirate signal processing that down-sampling a signal increases its frequency [25]. So by down-sampling the error, these low frequency oscillations can be converted to high frequency ones. Afterwards, the algorithm can be re-run to damp them out. When the error ceases to decrease, it can be further down-sampled and the algorithm is once again re-run. In this way, error can be reduced to any desirable tolerance in multiple cycles and at multiple scales. These multiple scales are known as multigrids and, hence, follows the name multigrid methods. Multigrid methods are the power house of boundary value problems arising in computational electromagnetics, computational fluid dynamics, statistical physics, wavelet theory, applied mathematics, and many other branches of computational science and engineering [18].

### 11.5. Preconditioning methods for large linear systems

Consider Eq. (6) for LMS algorithm once again,

$$x[k+1] = x[k] + \mu a_k(b_k - a_k{}^T x[k]) \tag{Repeat}$$

The term $(b_k - a_k{}^T x[k])$ is the residue for $k$-th row of $A_k$ matrix in Eq. (1). The residue is generated as soon as the $k$-th data row becomes available. This is ideal for run-time or on the fly operation. But if all the input and output data is available beforehand, say for example pre-recorded in a laboratory, then the entire $A_m$ matrix and the $b_m$ vector can be used at once in Eq. (6) instead of the individual data rows and observations.

$$x[k+1] = x[k] + \mu A^T{}_m(b_m - A_m x[k]) \tag{66}$$

We have used $A_m$ and $b_m$ instead of $A_k$ and $b_k$ because only $m$ input data rows and $m$ output observations are available in the lab. So, we have to find $x$ that agrees with all the available observations. No new measurements will be available. Now the question arises about the choice of best step-size for Eq. (66). If we want to solve Eq. (66) in one-step, the ideal step-size would be $\mu = (A^T{}_m A_m)^{-1}$.

$$x[k+1] = x[k] + (A^T{}_m A_m)^{-1} A^T{}_m(b_m - A_m x[k]) \tag{67}$$

Expanding Eq. (67),

$$x[k+1] = x[k] + (A^T{}_m A_m)^{-1} A^T{}_m b_m - (A^T{}_m A_m)^{-1} A^T{}_m A_m x[k] \qquad (68)$$

Finally,

$$x[k+1] = x[k] + x - x[k] = x \qquad (69)$$

So the system will converge in one step. Observe that for this to happen we practically need to multiply the residue term $(b_m - A_m x[k])$ in Eq. (67) with the pseudoinverse $(A^T{}_m A_m)^{-1} A^T{}_m$. But if the pseudinverse $(A^T{}_m A_m)^{-1} A^T{}_m$ is available beforehand, whole point of iterative solution becomes moot. Instead the system can be solved directly. Here the preconditioning methods come to rescue. These methods suggest that instead of multiplying the pseudoinverse $(A^T{}_m A_m)^{-1} A^T{}_m$ with the residue term, we can choose another matrix $P$ that is much easier to invert.

$$x[n+1] = x[n] + P^{-1}(b_m - A_m x[n]) \qquad (70)$$

$P$ can be the diagonal part of the pseudoinverse or may constitute its lower-triangular portions. Former choice is known as the *Jacobi iteration* and the latter one as the *Gauss-Seidel iteration*. *Successive Over Relaxation (SOR)* is a combination of both [18]. Anyhow, the idea is to find a $P$ that is much easier to invert. Solving Eq. (70),

$$P x[n+1] = P x[n] + (b_m - A_m x[n]) \qquad (71)$$

Or,

$$P x[n+1] = (P - A_m) x[n] + b_m \qquad (72)$$

Eq. (72) is known as the *preconditioning equation*. $P$ is known as the *preconditioning matrix*. The process $(P - A)$ is termed as *splitting*. Preconditioning methods are popular for solving large linear systems that are too expensive for traditional methods based on matrix factorizations.

### 11.6. Krylov subspace methods and conjugate gradients for optimization and control

If identity matrix is chosen as the precondition matrix $P$ and $b_m$ as the initial guess $x[n]$, Eq. (72) becomes,

$$x[n+1] = (I - A_m) b_m + b_m = 2 b_m - A_m b_m \qquad (73)$$

Iterating Eq. (73) for the second time,

$$x[n+2] = (I - A_m) x[n+1] + b_m = 3 b_m - 3 A_m b_m + A^2{}_m b_m \qquad (74)$$

$b_m, A_m b_m, \ldots, A^n{}_m b_m$ constitute the basis vectors for the Krylov subspace. According to Eq. (74), $x$ can be obtained by a linear combination of Krylov basis vectors. Hence, the solution to Eq. (1) can be found in Krylov subspace. This is the driving concept behind Krylov subspace methods. Advanced Krylov subspace methods include conjugate gradient method and minimum Residual methods: MINRES and GMRES [18]. These methods are central to optimization theory and automatic control systems.

## 12. DISCUSSION

Various authors have tried to address the problems mentioned in the Introduction section, though in part only. Widrow & Stearn [26], for example, have tried to present a broad conceptual introduction to the basic algorithms. Their treatment of subject is light on mathematics and is definitely not for someone who wishes to implement these algorithms. Boroujeny [1] has attempted to make these algorithms more implementable by providing a concrete mathematical approach. However, the usual clarity and the knack of exposition of the author just disappear as the algorithms become more advanced. For example, author just restates the major results in RLS filtering in terms of his own notation and drops the Kalman filter entirely. Poularikas [27] has made an effort to make the basic algorithms accessible by using the simulation approach. Yet again, Kalman filter is entirely left out and RLS is barely touched. Diniz [28] has provided a detailed treatment of these algorithms by following a statistical/linear algebraic approach to the subject. But when it comes to Kalman filter, the author switches to state space paradigm of control theory for a brief introduction to the topic. Strang [20, 21] has tried to tackle the RLS and Kalman filtering problem and has endeavored to bring it to linear algebraic perspective. But the author entirely leaves out the Wiener filters, Kaczmarz, LMS, and NLMS algorithms. Dwight [29] has tried to categorize the basic algorithms in two different paradigms, the least squares paradigm and MMSE paradigm using linear algebraic approach. But when it comes to the exposition of Kalman filter, the author completely switches to Wiener's system theory and statistical signal processing paradigm resulting in almost no connection between his earlier and later results. To unify adaptive signal processing and statistical signal processing platforms, an effort has been made by Manolakis et el [30]. But the book has rather obscure notation and the topics are almost inaccessible without familiarization with it. Chong [16] provides a treatment of only RLS and Kaczmarz algorithms from optimization theoretic viewpoint. Whereas Hayes [31], Syed [32], Haykin [33], and Candy [24] make no attempt at all at making these algorithms accessible.

As we have been through this journey, we have discovered two things that have shaped our methodology. Firstly, no matter how hidden it is behind all the veils just mentioned, it is all linear algebra. Secondly, from the basic Steepest Descent algorithm to fairly advanced Kalman filter, one and only one equation is involved. This equation assumes a different name and wears a different mantle of symbols in each algorithm. If somehow this link can be discovered, a relationship can be established among all the algorithms. Then the transition from one algorithm to the other will be more logical, more systematic, and much easier to understand. This will provide an opportunity to focus more on the underlying problem, the reason for the progression from one algorithm to the other, and the advantage provided by the new algorithm over the previous one. Also their notation can be made more uniform which will generate certain terseness in their explanation. In this way, one approach can be picked and followed to the end without the need for switching between different subjects. We have adopted a linear algebraic approach for this purpose. It is the simplest possible approach in a sense that it asks for no technical introduction, making it possible for everyone to benefit. We have started from the simplest possible case, one equation in one unknown and one observation. From this one equation, we have derived all the key algorithms from SD to Kalman filter while leaving nothing out. Notation was kept consistent throughout. Transitions from one algorithm to the other were logical. They were also systematic to make the concepts portable. Calculus, Probability, and Statistics were not invoked in order to make them accessible as well. The treatment was entirely linear algebraic. Each step was explained but the explanations were kept concise in order to prevent the manuscript from becoming too voluminous.

## 13. CONCLUSION

Our work provides a uniform approach to students, researchers, and practitioners from different academic backgrounds working in different areas. Anyone can learn these key algorithms without any previous knowledge. The understanding of readers will be enhanced and their efforts will be minimized as they will see just one equation in action. It will enable them to focus more on the concept rather than the symbols, helping them to master the essentials in minimum amount of time. As there is only one equation to deal with, it will also ease their burden of programming. They can program it without the need of special programming skills. They will be able to appreciate how just one equation has been the centre of research and development for last many decades and how it has vexed scientists and engineers in many different fields. In fact, to demonstrate the benefit of this synthesis, an entire area in applied mathematics known as multigirid methods that aim to solve differential equations by operating at multiple resolutions, stems from this one equation. Other specialized domains like non-linear control and particle filter theory also take the lead from this one equation. Finally, a better understanding of these algorithms will ultimately lead to a better understanding of image processing, video processing, wireless communications, pattern recognition, machine learning, optimization theory and many other subjects that utilize these core concepts. For anyone, this one equation will eventually be their passport to the realm of computational science and engineering. For students, this one equation will save them the toil of going through these algorithms again and again in various courses where they just change their name and switch their clothes. For faculty, this one equation will spare their efforts to teach these algorithms repeatedly in various courses. For universities, this one equation will enable them to develop a consolidated course based on these algorithms that are central to so many subjects in the ECE and applied mathematics.

## REFERENCES

[1]     B. Farhang-Boroujeny, *Adaptive Filters: Theory and Applications*: John Wiley & Sons, 2013.
[2]     P. Avirajamanjula and P. Palanivel, "Corroboration of Normalized Least Mean Square Based Adaptive Selective Current Harmonic Elimination in Voltage Source Inverter using DSP Processor," *International Journal of Power Electronics and Drive Systems (IJPEDS)*, vol. 6, 2015.

[3]     Y. Zheng-Hua, R. Zi-Hui, Z. Xian-Hua, and L. Shi-Chun, "Path Planning for Coalmine Rescue Robot Based on Hybrid Adaptive Artificial Fish Swarm Algorithm," *TELKOMNIKA Indonesian Journal of Electrical Engineering,* vol. 12, pp. 7223-7232, 2014.

[4]     F. Wang and Z. Zhang, "An Adaptive Genetic Algorithm for Mesh Based NoC Application Mapping," *TELKOMNIKA Indonesian Journal of Electrical Engineering,* vol. 12, pp. 7869-7875, 2014.

[5]     X. Yan, J. Li, Z. Li, Y. Yang, and C. Zhai, "Design of Adaptive Filter for Laser Gyro," *TELKOMNIKA Indonesian Journal of Electrical Engineering,* vol. 12, pp. 7816-7823, 2014.

[6]     J. Mohammed, "Low Complexity Adaptive Noise Canceller for Mobile Phones Based Remote Health Monitoring," *International Journal of Electrical and Computer Engineering (IJECE),* vol. 4, pp. 422-432, 2014.

[7]     K. F. Akingbade and I. A. Alimi, "Separation of Digital Audio Signals using Least Mean Square LMS Adaptive Algorithm," *International Journal of Electrical and Computer Engineering (IJECE),* vol. 4, pp. 557-560, 2014.

[8]     J. Mohammed and M. Shafi, "An Efficient Adaptive Noise Cancellation Scheme Using ALE and NLMS Filters," *International Journal of Electrical and Computer Engineering (IJECE),* vol. 2, pp. 325-332, 2012.

[9]     J. Mohammed, "A Study on the Suitability of Genetic Algorithm for Adaptive Channel Equalization," *International Journal of Electrical and Computer Engineering (IJECE),* vol. 2, pp. 285-292, 2012.

[10]    J. Huang, J. Xie, H. Li, G. Tian, and X. Chen, "Self Adaptive Decomposition Level Denoising Method Based on Wavelet Transform," *TELKOMNIKA Indonesian Journal of Electrical Engineering,* vol. 10, pp. 1015-1020, 2012.

[11]    N. Makwana, M. S.P.I.T, N. Mishra, and B. Sagar, "Hilbert Transform Based Adaptive ECG R Peak Detection Technique," *International Journal of Electrical and Computer Engineering (IJECE),* vol. 2, pp. 639-643, 2012.

[12]    Z. Chen, U. Qiqihar, X. Dai, L. Jiang, C. Yang, and B. Cai, "Adaptive Iterated Square Root Cubature Kalman Filter and Its Application to SLAM of a Mobile Robot," *TELKOMNIKA Indonesian Journal of Electrical Engineering,* vol. 11, pp. 7213-7221, 2013.

[13]    Z. Ji, "Adaptive Cancellation of Light Relative Intensity Noise for Fiber Optic Gyroscope," *TELKOMNIKA Indonesian Journal of Electrical Engineering,* vol. 11, pp. 7490-7499, 2013.

[14]    Q. Liu, "An Adaptive Blind Watermarking Algorithm for Color Image," *TELKOMNIKA Indonesian Journal of Electrical Engineering,* vol. 11, pp. 302-309, 2013.

[15]    G. Strang, "Introduction to Linear Algebra," *Wellesley-Cambridge Press,* 2003.

[16]    E. K. Chong and S. H. Zak, *An Introduction to Optimization* vol. 76: John Wiley & Sons, 2013.

[17]    K.-A. Lee, W.-S. Gan, and S. M. Kuo, *Subband Adaptive Filtering: Theory and Implementation*: John Wiley & Sons, 2009.

[18]    G. Strang, *Computational Science and Engineering*: Wellesley-Cambridge Press, 2007.

[19]    T. S. Rappaport, *Wireless Communications: Principles and Practice* vol. 2: prentice hall PTR New Jersey, 1996.

[20]    G. Strang and K. Borre, *Linear Algebra, Geodesy, and GPS*: Wellesley-Cambrdige Press, 1997.

[21]    G. Strang, *Introduction to Applied Mathematics* vol. 16: Wellesley-Cambridge Press, 1986.

[22]    J. S. Goldstein and I. S. Reed, "Reduced-rank Adaptive Filtering," *IEEE Transactions on Signal Processing,* vol. 45, pp. 492-496, 1997.

[23]    J. Mochnác, S. Marchevský, and P. Kocan, "Bayesian Filtering Techniques: Kalman and Extended Kalman Filter Basics," in *Radioelektronika, 2009. RADIOELEKTRONIKA'09. 19th International Conference,* 2009, pp. 119-122.

[24]    J. V. Candy, *Bayesian Signal Processing: Classical, Modern and Particle Filtering Methods* vol. 54: John Wiley & Sons, 2011.

[25]    G. Strang and T. Nguyen, *Wavelets and Filter Banks*: Wellesley-Cambrdige Press, 1996.

[26]    B. Widrow and S. D. Stearns, *Adaptive Signal Processing* vol. 1: Prentice-Hall, Inc., 1985.

[27]    A. D. Poularikas and Z. M. Ramadan, *Adaptive Filtering Primer with MATLAB*: CRC Press, 2006.

[28]    P. S. Diniz, *Adaptive Filtering*: Springer, 1997.

[29]    D. F. Mix, *Random Signal Processing*: Prentice-Hall, Inc., 1995.

[30]    D. G. Manolakis, V. K. Ingle, and S. M. Kogon, *Statistical and Adaptive Signal Processing: Spectral Estimation, Signal Modeling, Adaptive Filtering, and Array Processing* vol. 46: Artech House Norwood, 2005.

[31]    M. H. Hayes, *Statistical Digital Signal Processing and Modeling*: John Wiley & Sons, 1996.

[32]    A. H. Sayed, *Fundamentals of Adaptive Filtering*: John Wiley & Sons, 2003.

[33]    S. S. Haykin, *Adaptive Filter Theory*: Pearson Education India, 2008.